\documentclass[aps, prb, superscriptaddress, twocolumn]{revtex4}
\usepackage{amssymb}
\usepackage{graphicx}
\usepackage{hyperref}

\begin{document}

\title{Magnetic, thermal, and transport properties of the mixed valent vanadium oxides LuV$_4$O$_8$ and YV$_4$O$_8$}
\author{S. Das}
\affiliation{Ames Laboratory and Department of Physics and Astronomy, Iowa State University, Ames, Iowa 50011}
\author{A. Niazi}
\affiliation{Ames Laboratory and Department of Physics and Astronomy, Iowa State University, Ames, Iowa 50011}
\author{Y. Mudryk}
\affiliation{Ames Laboratory, Iowa State University, Ames, Iowa 50011}
\author{V. K. Pecharsky}
\affiliation{Ames Laboratory and Department of Materials Science and Engineering, Iowa State University, Ames, Iowa 50011}
\author{D. C. Johnston}
\affiliation{Ames Laboratory and Department of Physics and Astronomy, Iowa State University, Ames, Iowa 50011}

\date{\today}
    
\begin{abstract}

\textit{L}V$_4$O$_8$ (\textit{L} = Yb, Y, Lu) compounds are reported to crystallize in a structure similar to that of the orthorhombic CaFe$_2$O$_4$ structure-type, and contain four inequivalent V sites arranged in zigzag chains. We confirm the structure and report the magnetic, thermal, and transport properties of polycrystalline ${\rm YV_4O_8}$ and ${\rm LuV_4O_8}$. A first-order like phase transition is observed at 50 K in both YV$_4$O$_8$ and LuV$_4$O$_8$. The symmetry remains the same with the lattice parameters changing discontinously. The structural transition in YV$_4$O$_8$ leads to partial dimerization of the V atoms resulting in a sudden sharp drop in the magnetic susceptibility. The V spins that do not form dimers order in a canted antiferromagnetic state. The magnetic susceptibility of LuV$_4$O$_8$ shows a sharp peak at $\sim 50$ K\@. The magnetic entropies calculated from heat capacity versus temperature measurements indicate bulk magnetic transitions below 90 K for both YV$_4$O$_8$ and LuV$_4$O$_8$.

\end{abstract}
\maketitle

\section{\label{intro}Introduction}

\begin{figure}[t]
\includegraphics[width=3in]{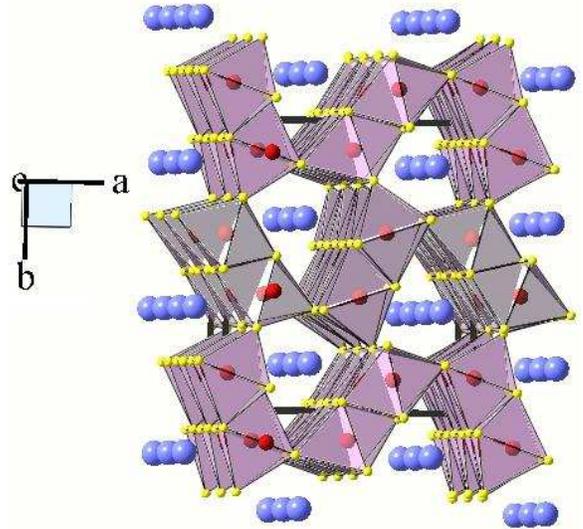}
\caption{(Color online) Crystal structure of the low-temperature $\alpha$-phase of \textit{L}V$_4$O$_8$ viewed along the $c$-axis. The large blue, intermediate red, and small yellow circles represent $L$, V and O atoms, respectively. The VO$_6$ octahedra share edges to form V zigzag chains running along the $c$-axis. The $L$ ions occupy half of the cation sites in the ${\rm CaV_2O_4}$ structure in an ordered fashion while the other half is vacant.}
\label{struc_c}
\end{figure}

\begin{figure}[t]
\includegraphics[width=3in]{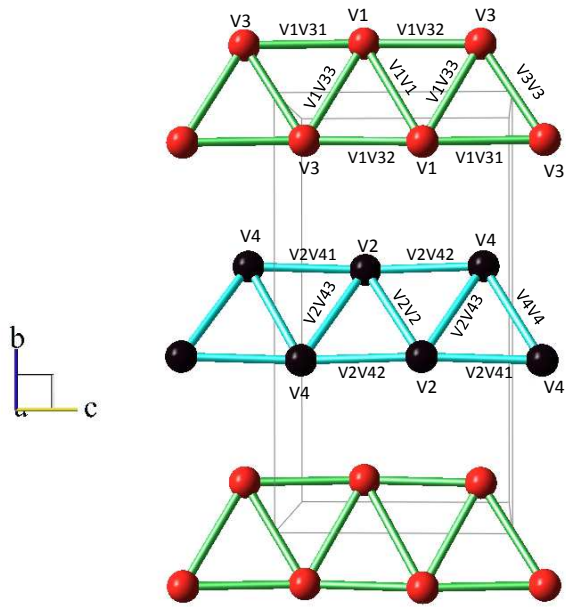}
\caption{(Color online) The arrangment of V atoms in the structure as viewed along the $a$-axis. Two distinct chains are formed by V atoms in four inequivalent sites labelled 
V1, V2, V3, and V4, respectively. The bond lengths between different V atoms are labelled as V1V1, V2V2, V3V3, V4V4, V1V31, V1V32, V1V33, V2V41, V2V42, and V2V43, respectively.}
\label{VVdisstruc}
\end{figure}

Vanadium oxides have been of broad interest owing to their interesting properties. Binary vanadium oxides V$_n$O$_{2n-1}$ where 2 $\leq n \leq 9$ exhibit metal to insulator and paramagnetic to antiferromagnetic transitions on cooling.\cite{udo2004} The only exception is V$_7$O$_{13}$ which remains metallic down to 4 K\@.\cite{kachi73} Among ternary vanadium oxides, the normal spinel mixed valent ${\rm LiV_2O_4}$ does not show any magnetic ordering, remains metallic down to 0.5 K and surprisingly shows heavy fermion behavior below 10 K\@.\cite{kondo97} This is very different from the similar normal spinel ${\rm LiTi_2O_4}$ which shows superconductivity below 13 K\@.\cite{johnston77} 

The compound ${\rm CaV_2O_4}$ forms in the well-known ${\rm CaFe_2O_4}$ type structure with orthorhombic space group $Pnam$ and lattice parameters $a = 9.206$~\AA, $b = 10.674$~\AA, and $c = 3.009$~\AA.\cite{decker,hastings,niazi2009} The V atoms have spin $S=1$ and form a zigzag chain system. The compound undergoes an orthorhombic to monoclinic structural distortion below 150 K and an antiferromagnetic transition at 63 K, and is an insulator.\cite{zong2008,niazi2009,pieper2009} The low dimensionality of the V spin structure is very interesting since this can give rise to exotic magnetism. Indeed, there is a suggestion that a phase transition at $\simeq 200$~K in CaV$_2$O$_4$ arises from a long- to short-range chiral ordering transition with no long-range spin order either below or above this temperature.\cite{niazi2009} In a spin $S=1$ zigzag chain system, depending on the ratio of the nearest-neighbor and next-nearest-neighbor interactions, there can be ground states with a Haldane gap, as well as gapless or gapped chiral ordering.\cite{hikihara} Replacing Ca$^{2+}$ by Na$^{+1}$, the same ${\rm CaFe_2O_4}$ structure is retained but becomes metallic even below the antiferromagnetic transition at 140 K\@.\cite{Yamaura2007,Sakurai2008} Further investigations of compounds having the ${\rm CaFe_2O_4}$-type and related structures are clearly warranted. 

The compounds \textit{L}V$_4$O$_8$ ($L$ = Yb, Y, Lu) have structures similar to the ${\rm CaFe_2O_4}$-type structure but with the modification that in \textit{L}V$_4$O$_8$, only half of the $L$ cation sites are occupied by $L$ ions in an ordered manner.\cite{kanke97} This results in a reduction of the unit cell symmetry from orthorhombic to monoclinic with space group $P12_1$/$n1$ (which is a nonisomorphic subgroup of the orthorhombic space group $Pnam$ of CaV$_2$O$_4$) and lattice parameters $a = 9.0648$ \AA, $b = 10.6215$ \AA, $c = 5.7607$ \AA, and $\beta = 90.184^\circ$ for the room temperature $\alpha$-phase (see below) of the Yb compound.\cite{kanke97} Note that the monoclinic angle $\beta$ is close to 90$^\circ$ and that the $a$-axis and $b$-axis lattice parameters are nearly the same as in the above orthorhombic room-temperature structure of CaV$_2$O$_4$. Figure~\ref{struc_c} shows the structure of $\alpha$-\textit{L}V$_4O_8$ viewed along the $c$ axis. The slightly distorted VO$_6$ octahedra share edges and corners to form zigzag chains along the $c$ axis. The four V atoms in the structure occupy four inequivalent positions and form two distinct chains with two inequivalent V positions in each chain. The V-V zigzag chains as viewed along the $a$ axis are shown in Fig.~\ref{VVdisstruc}. 

YbV$_4$O$_8$ forms in two monoclinic phases, the low temperature $\alpha$-phase with space group $P12_1$/$n1$ and lattice parameters $a = 9.0648$~\AA, $b = 10.6215$~\AA, $c = 5.7607$~\AA, and $\beta = 90.184^\circ$ and the high temperature $\beta$-phase with space group $P2_1/n11$ and lattice parameters $a = 9.0625$\AA, $b = 11.0086$~\AA, $c = 5.7655$~\AA, and $\alpha = 105.070^\circ$.\cite{kanke97} The two phases differ crystallographically by the $z$ atomic position of the Yb ions, but both contain similar zigzag chains. At 185 K the $\beta$-${\rm YbV_4O_8}$ undergoes a magnetic phase transition with magnetic behavior of the vanadium cations separating into Curie-Weiss and spin gap types. The magnetic transition is accompanied at the same temperature by a monoclinic to monoclinic structural phase transition arising from complete charge ordering of the V$^{+3}$ and V$^{+4}$ ions.\cite{friese07} ${\rm YV_4O_8}$ also cystallizes in $\alpha$ and $\beta$ forms isomorphous with $\alpha$- and $ \beta$-${\rm YbV_4O_8}$.\cite{onoda03} ${\rm LuV_4O_8}$ was reported to have a homogeneity range from ${\rm LuV_4O_{7.93}}$ to ${\rm LuV_4O_{8.05}}$\cite{kitayama78} and its structure is isostructural with $\alpha$-YbV$_4$O$_8$.\cite{kanke97}

The structures of the above \textit{L}V$_4$O$_8$ compounds are closely related to the Hollandite-type structure with either tetragonal or monoclinic crystal symmetry and chemical formula $A_xB_8$O$_{16}$ ($A$ = K, Li, Sr, Ba, Bi; $B$ = Ti, V, Mn, Ru, Rh; $1 \leq x \leq 2$).\cite{bystrom,torardi1985} In the Hollandites, edge-sharing $B$O$_6$ octahedra form zigzag chains running parallel to the crystallographic $c$ axis. The Hollandite K$_2$V$_8$O$_{16}$ undergoes a metal-isulator and a structural transition at 170~K which leads to possible dimerization of the V spins.\cite{isobe} The presence of a quantum phase transition from a weakly localized state to a metallic state in BaRu$_6$O$_{12}$ has been reported.\cite{cava2003}

The magnetic susceptibilties of $\alpha$-${\rm YV_4O_8}$ and $\beta$-${\rm YV_4O_8}$ show Curie-Weiss behavior in the high \textit{T} region and drop sharply on cooling to temperatures between 50 and 80 K\@.\cite{onoda03} For $\alpha$-${\rm YV_4O_8}$, the drop at 50 K appears to be a first order transition. This is different from the magnetic susceptibility of the isostructural ${\rm YbV_4O_8}$ or similiarly structured ${\rm CaV_2O_4}$.\cite{zong2008} Curie-Weiss fits to the high $T$ susceptibilities yielded negative Weiss temperatures indicating dominant antiferromagnetic interactions among the V spins and Curie constants much lower than expected for three V$^{+3}$ ($S = 1$) and one V$^{+4}$ ($S = 1/2$) spins per formula unit for both $\alpha$- and $\beta$-YV$_4$O$_8$. In order to investigate the origin of the first order-like transition in YV$_4$O$_8$ and to search for interesting magnetic ground states in these zigzag spin chain systems with modified CaFe$_2$O$_4$ crystallographic structure, we have synthesized polycrystalline samples of YV$_4$O$_8$ and LuV$_4$O$_8$ and report their structure, magnetic susceptibility $\chi$, magnetization $M$, specific heat $C$, and the electrical resistivity $\rho$. 

The remainder of the paper is organized as follows. In Sec.~\ref{expt}, the synthesis procedure  and other experimental details are reported. The structures from room temperature down to 10 K, magnetic susceptibility, magnetization, heat capacity, and electrical resistivity measurements are presented in Sec.~\ref{results}. We also carried out bond valence analysis to estimate the valences of the inequivalent V atoms in the mixed valent YV$_4$O$_8$ and LuV$_4$O$_8$ compounds. The results of this analysis are reported following the x-ray diffraction measurements in Sec.~\ref{results}. In Sec.~\ref{discussion}, we suggest a model to explain the observed magnetic susceptibility and heat capacity behaviors of YV$_4$O$_8$ in light of the structural studies reported in Sec.~\ref{results}, whereas a model to explain the magnetic susceptibility and heat capacity behaviors of LuV$_4$O$_8$ is elusive. A summary of our results is given in Sec.~\ref{summary}.

\section{\label{expt}Experimental Details}

The samples of \textit{L}V$_4$O$_8$ ($L$ = Y, Lu) were prepared by solid state reaction. The starting materials for our samples were ${\rm Y_2O_3}$ (99.995\%, Alfa Aesar), ${\rm Lu_2O_3}$, ${\rm V_2O_5}$ (99.999\%, MV Laboratories Inc.), and ${\rm V_2O_3}$ (99.999\%, MV Laboratories Inc.). Stoichiometeric amounts of \textit{L}$_2$O$_3$,${\rm V_2O_5}$, and ${\rm V_2O_3}$ were thoroughly mixed together in a glove box filled with helium gas, and pressed into pellets. The pellets were wrapped in platinum foils, sealed in evacuated quartz tubes and heated at 520~$^\circ$C for 8--10 d. The temperature was then raised to 800~$^\circ$C for another 5--7 d. Finally the samples were heated at 1200 $^\circ$C for another 7 d. The quartz tubes were then taken out of the furnace at 1200~$^\circ$C and quenched in air to room temperature. 

Powder x-ray diffraction measurements at room temperature were done using a Rigaku Geigerflex diffractometer with a curved graphite crystal monochromator. Temperature-dependent powder x-ray diffraction studies were done in the temperature range 10~K -- 295~K using a standard Rigaku TTRAX diffractometer system equipped with a theta/theta wide-angle goniometer and a Mo $K\alpha$ radiation source.\cite{holm} The magnetic measurements were done using a Quantum Design superconducting quantum interference device (SQUID) magnetometer in the temperature range 1.8~K -- 350~K and magnetic field range 0 -- 5.5~T\@. The heat capacity and electrical resistivity measurements were done using a Quantum Design physical property measurement system (PPMS). For the heat capacity measurements, Apiezon N grease was used for thermal coupling between the samples and the sample platform. Heat capacity was measured in the temperature range 1.8 K -- 320~K in zero, 5 T, and 9 T magnetic fields. Electrical resistivity measurements were carried out using a standard dc 4-probe technique. Platinum leads were attached to rectangular shaped pieces of sintered pellets using silver epoxy. An excitation current of 10~mA was used in the resistivity measurements in the temperature range 1.8~K -- 300 K\@.

\section{\label{results}Results}

\subsection{\label{xray}X-ray diffraction measurements}

\begin{figure}[t]
\includegraphics[width=3in]{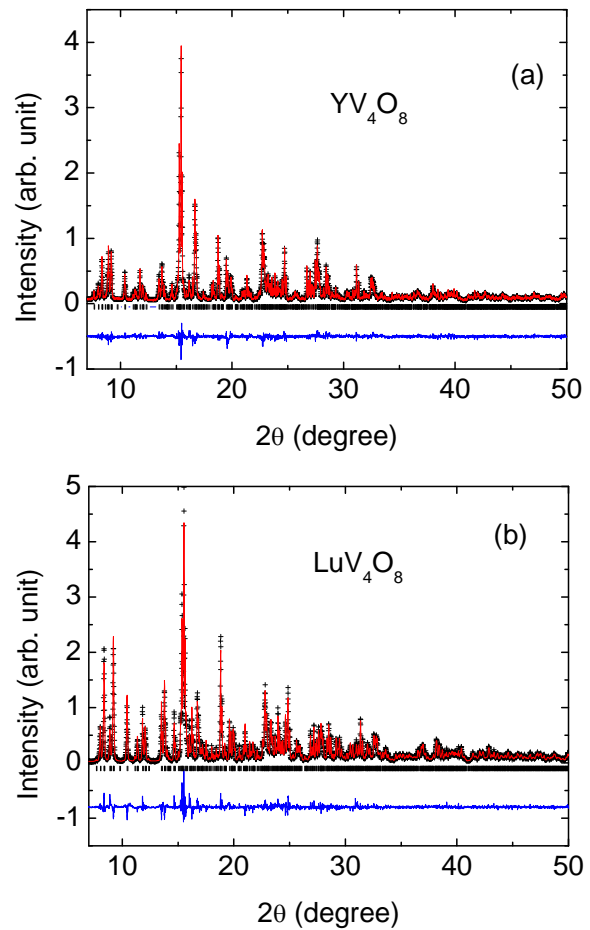}
\caption{(Color online) X-ray diffraction patterns of ${\rm YV_4O_8}$ (a) and ${\rm LuV_4O_8}$ (b), respectively, at room temperature. The solid crosses are the observed data points while the solid lines are the Rietveld fits to the data. The tic marks below the data indicate the peak positions. The solid lines below the tick marks are the difference between the observed and the calculated intensities. Small amounts ($<$~4~wt\%) of V$_2$O$_3$ impurity phases are present in both ${\rm YV_4O_8}$ and ${\rm LuV_4O_8}$ samples.}
\label{refine}
\end{figure}

\begin{table}
\caption{Lattice parameters and the fractional atomic positions of ${\rm YV_4O_8}$ at 295 K, obtained by Rietveld refinement of powder XRD data. Space group: $P12_1$/$n1$ (No.~14); $Z = 4$ formula units/unit cell; lattice parameters: $a$~=~9.1186(2)~\AA, $b$~=~10.6775(2)~\AA, $c$~=~5.7764(1)~\AA, and monoclinic angle $\beta = 90.206(1)^\circ$; $R(F^2) = 0.083$. All atoms are in general Wyckoff positions 4($e$): $x$, $y$, $z$\@. A number in parentheses gives the error in the last or last two digits of the respective quantity.}
\begin{ruledtabular}
\begin{tabular}{llll}
 & $x$ & $y$ & $z$\\
\hline
Y1 & 0.7574(2) & 0.6581(2) & 0.1257(4)\\
V1 & 0.4282(3) & 0.6175(3) & 0.1266(8))\\
V2 & 0.4107(3) & 0.0989(3) & 0.1235(9)\\
V3 & 0.4537(3) & 0.6111(3) & 0.6263(8)\\
V4 & 0.4193(3) & 0.1043(3) & 0.6252(9)\\
O1 & 0.1977(9)  & 0.1516(1) & 0.0977(21)\\
O2 & 0.1154(9) & 0.4760(10) & 0.1266(29)\\
O3 & 0.5278(9) & 0.7744(9) & 0.1285(30)\\
O4 & 0.4238(11) & 0.4297(9) & 0.1177(33)\\
O5 & 0.2198(9) & 0.1492(10) & 0.6164(22)\\
O6 & 0.1195(10) & 0.4800(11) & 0.6227(27)\\
O7 & 0.5119(10) & 0.7934(9) & 0.6155(28)\\
O8 & 0.4130(11) & 0.4287(9) & 0.6450(30)\\
\end{tabular}
\end{ruledtabular}
\label{tableY}
\end{table}

\begin{table}
\caption{Lattice parameters and the fractional atomic positions of ${\rm LuV_4O_8}$ at 295 K, obtained by Reitveld refinement of powder XRD data. Space group: $P12_1$/$n1$ (No.~14); $Z = 4$ formula units/unit cell; lattice parameters: $a$~=~9.0598(2)~\AA, $b$~=~10.6158(2)~\AA, $c$~=~5.7637(1)~\AA, and monoclinic angle $\beta = 90.189(2)^\circ$; $R(F^2) = 0.095$. All atoms are in general Wyckoff positions 4($e$): $x$, $y$, $z$\@. A number in parentheses gives the error in the last or last two digits of the respective quantity.}
\begin{ruledtabular}
\begin{tabular}{llll}

& $x$ & $y$ & $z$\\
\hline
Lu1 & 0.7573(2) & 0.6583(1) & 0.159(2)\\
V1 & 0.4269(4) & 0.6170(4) & 0.1281(11)\\
V2 & 0.4103(4) & 0.0976(4) & 0.1217(13)\\
V3 & 0.4549(4) & 0.6107(4) & 0.6332(11)\\
V4 & 0.4182(4) & 0.1046(4) & 0.6230(12)\\
O1 & 0.2019(13)  & 0.1609(13) & 0.1091(33)\\
O2 & 0.1250(15) & 0.4698(14) & 0.1278(42)\\
O3 & 0.5299(14) & 0.7774(14) & 0.1258(45)\\
O4 & 0.4158(16) & 0.4237(12) & 0.1341(42)\\
O5 & 0.2098(13) & 0.1670(12) & 0.6358(34)\\
O6 & 0.1221(16) & 0.4747(14) & 0.6311(41)\\
O7 & 0.5135(14) & 0.7938(14) & 0.6283(42)\\
O8 & 0.4095(16) & 0.4314(14) & 0.6382(43)\\
\end{tabular}
\end{ruledtabular}
\label{tableLu}
\end{table}

\begin{figure*}[t]
\includegraphics[width=4in]{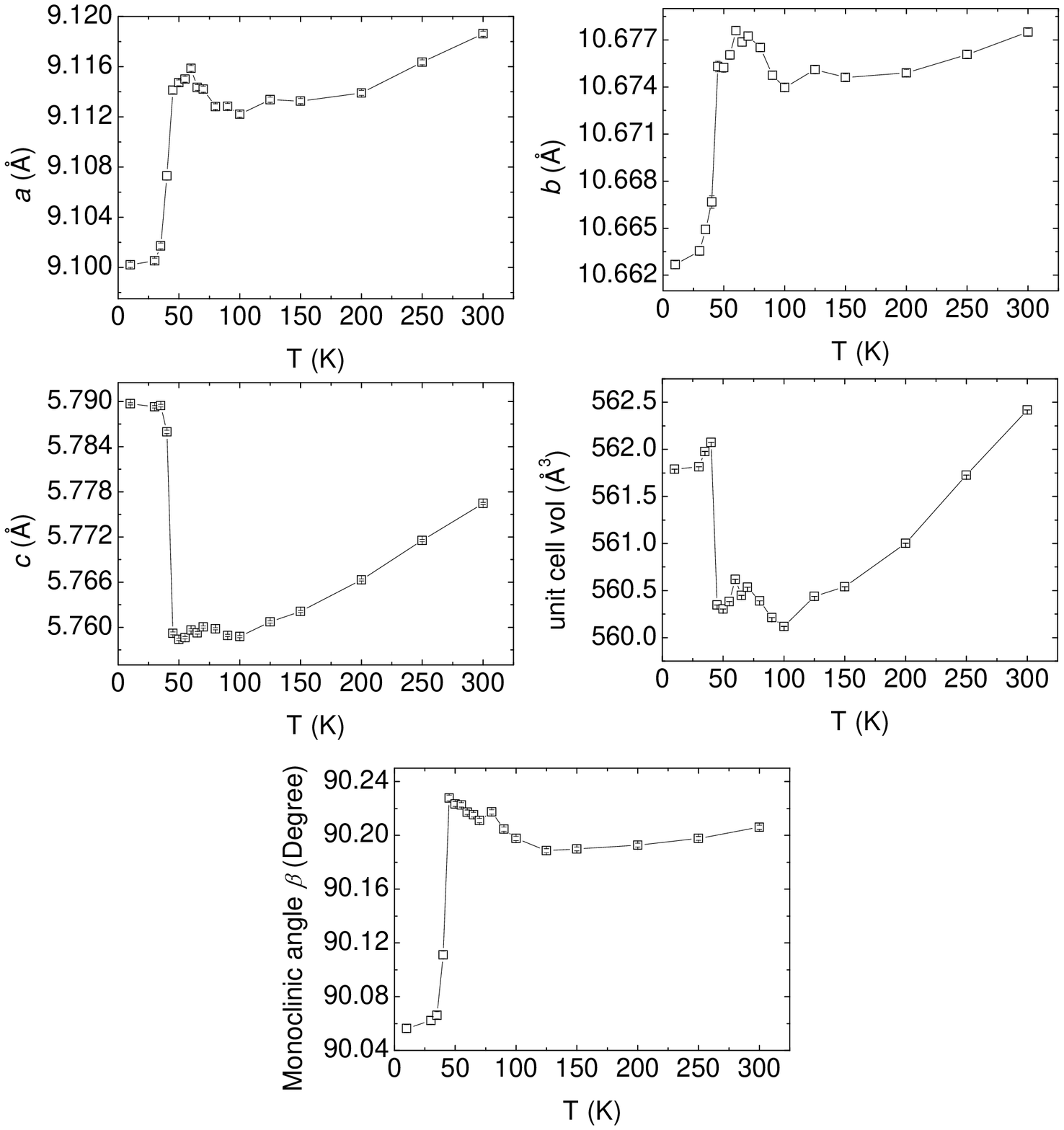}
\caption{Lattice parameters $a$, $b$, $c$, unit cell volume, and the monoclinic angle $\beta$ of ${\rm YV_4O_8}$ versus temperature $T$.}
\label{ylatall}
\end{figure*}

\begin{figure*}[t]
\includegraphics[width=4in]{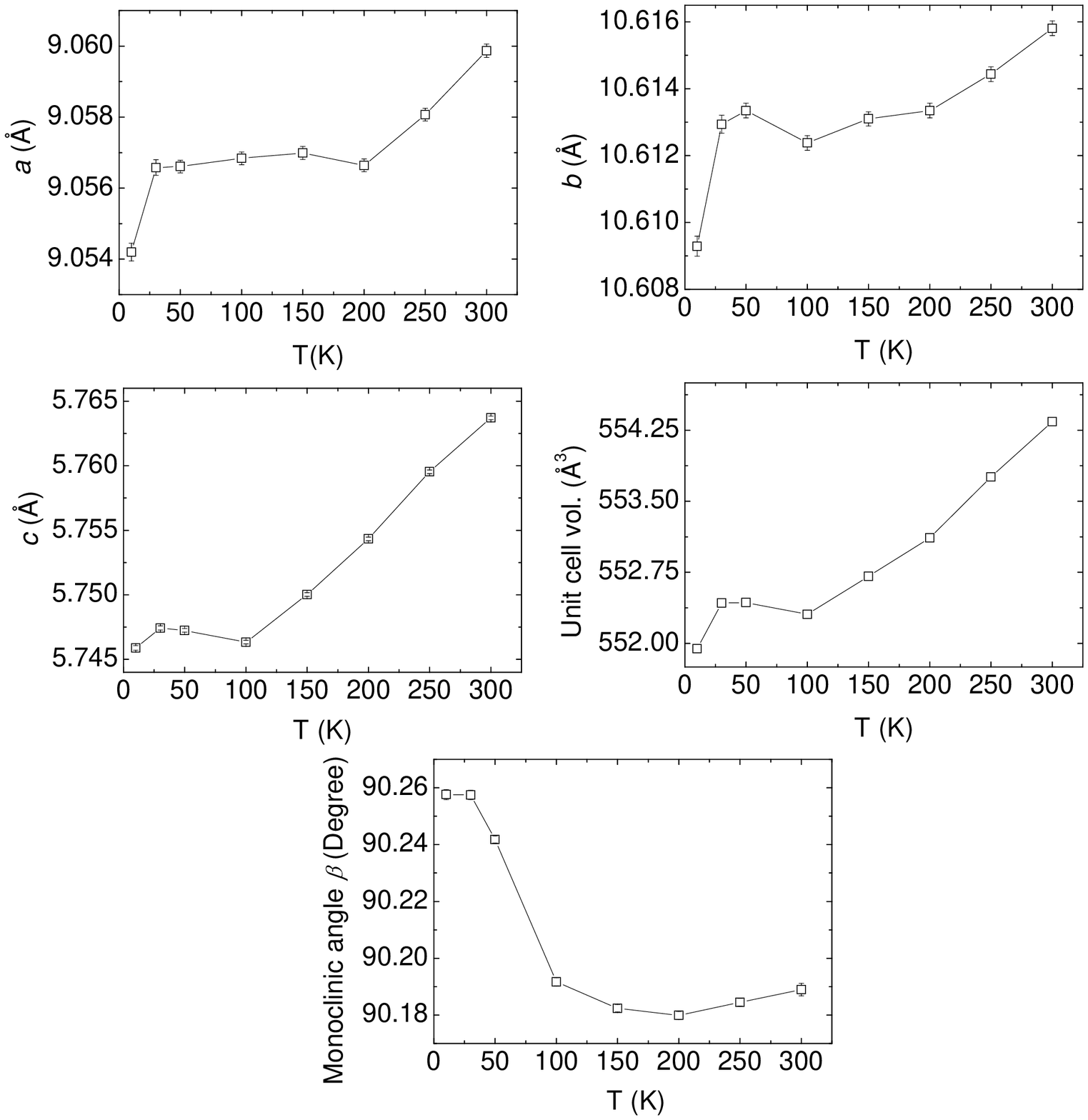}
\caption{Lattice parameters $a$, $b$, $c$, unit cell volume, and the monoclinic angle $\beta$ of ${\rm LuV_4O_8}$ versus temperature $T$.}
\label{lulatall}
\end{figure*}

\begin{figure*}[t]
\includegraphics[width=5in]{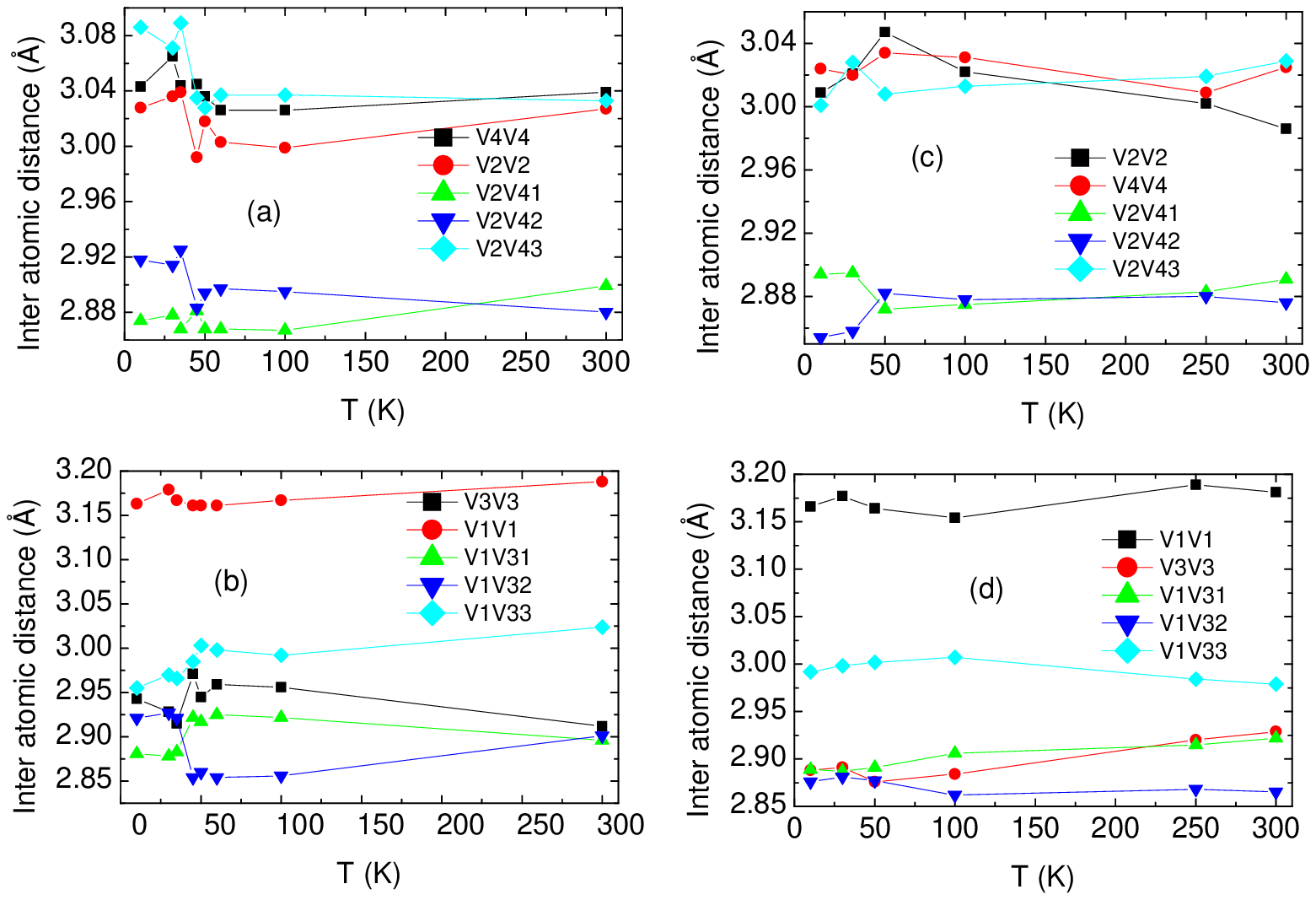}
\caption{(Color online) V-V bond lengths in (a)-(b) YV$_4$O$_8$ and (c)-(d) LuV$_4$O$_8$. For the atom notations see Fig.~\ref{VVdisstruc}.}
\label{VVdis}
\end{figure*}

Figures~\ref{refine}(a) and (b) show the room temperature x-ray diffraction (XRD) patterns of powder samples of ${\rm YV_4O_8}$ and ${\rm LuV_4O_8}$, respectively, along with the calculated patterns. The calculated patterns were obtained by Rietfeld refinements of the observed patterns using the GSAS program suite.\cite{gsas} The refinements for both ${\rm YV_4O_8}$ and ${\rm LuV_4O_8}$ were done with space group $P12_1$/$n1$ (No.~14) (the same space group as for the low-$T$ $\alpha$-phase of YbV$_4$O$_8$) with one position for the $L$ atom, four different positions for V atoms, and eight different positions for O atoms. All the fractional atomic positions, the lattice parameters, and the overall thermal parameter for all the atoms were varied in the refinement. The obtained best-fit lattice parameters and fractional atomic positions at 300 K are listed in Tables~\ref{tableY} and \ref{tableLu} for ${\rm YV_4O_8}$ and ${\rm LuV_4O_8}$, respectively. From the refinements, small amounts ($<$~4~wt\%) of V$_2$O$_3$ impurity phases were found in both ${\rm YV_4O_8}$ and ${\rm LuV_4O_8}$ samples.

Figure~\ref{ylatall} shows the lattice parameters $a$, $b$, $c$, unit cell volume, and the monoclinic angle $\beta$ respectively, of ${\rm YV_4O_8}$ versus temperature. At~$\sim$~50~K the $a$ and $b$ axes and the monoclinic angle $\alpha$ decrease sharply while the $c$ axis and the unit cell volume increase. There is no change in the symmetry of the unit cell. The sharp change in the lattice parameters and the unit cell volume indicate a first order phase transition.

For ${\rm LuV_4O_8}$, as shown in Fig.~\ref{lulatall}, the $a$ and $b$ lattice parameters decrease sharply below 45 K while the $c$ lattice parameter and the unit cell volume show a broad peak at $\sim 45$ K\@. The monoclinic angle $\beta$ increases below 100 K\@.

Figures~\ref{VVdis}(a)-(b) and \ref{VVdis}(c)-(d) show the V-V bond lengths versus temperature for different inequivalent V atoms in YV$_4$O$_8$ and LuV$_4$O$_8$, respectively. For both YV$_4$O$_8$ and LuV$_4$O$_8$, the V atoms at the four inequivalent sites form two different kinds of chains V1-V3 and V2-V4 running along the $c$ axis as shown in Fig.~\ref{VVdisstruc}. For the V1-V3 chain in YV$_4$O$_8$, the V1V32 distance increases while the V1V31 distance decreases below 50 K\@. The other V1-V3 distances also decrease below 50 K\@. For LuV$_4$O$_8$, the V2V42 distance decreases while the V2V41 distance increases below 50 K\@.

\subsubsection*{\label{BV}Bond Valence Analysis}

\begin{figure}[t]
\includegraphics[width=3in]{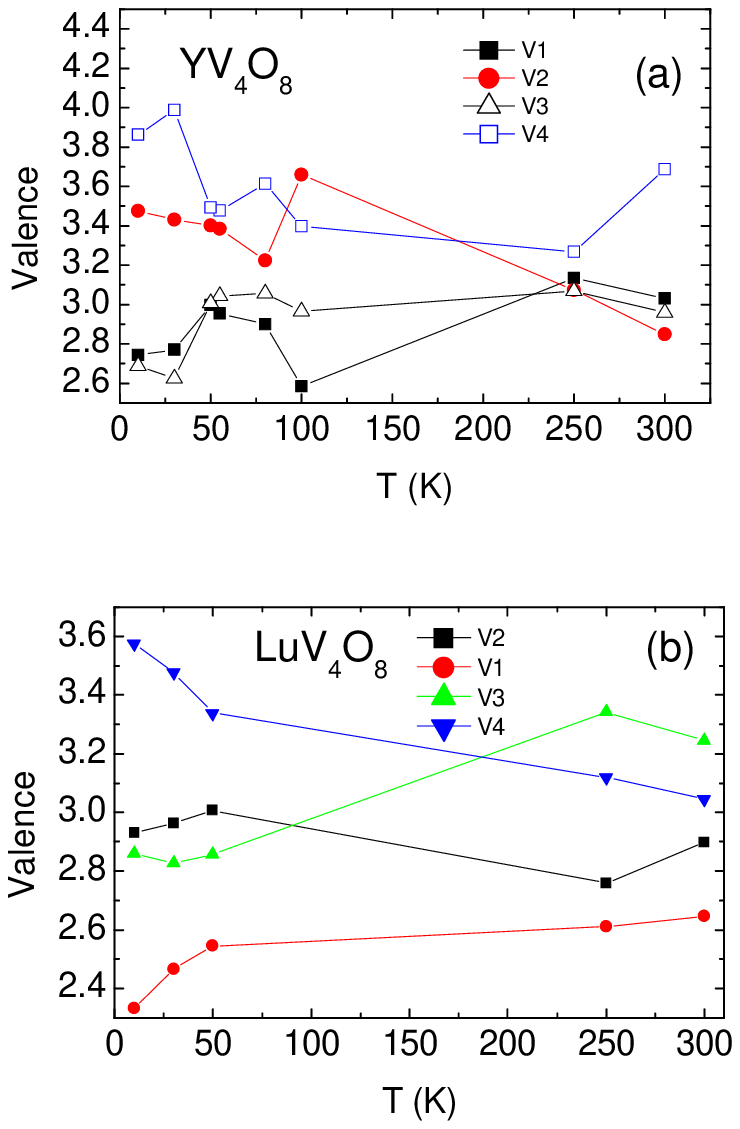}
\caption{(Color online) Valences of the different inequivalent V atoms versus temperature $T$ in (a) YV$_4$O$_8$ and (b) LuV$_4$O$_8$.}
\label{valency}
\end{figure}

The bond-valence method is used to calculate the valences of individual atoms in a chemical compound.\cite{Brown1973} The atomic valence of an atom is taken to be the sum of the bond valences of all bonds between that particular atom and the neighbouring atoms to which it is bonded. The bond-valence is defined as $v_i = \textrm{exp}[(r_0-r_i)/B]$ where $B$ is fixed to the value 0.37, $r_i$ is the interatomic distance between the particular atom and the neighbouring atom it is bonded to and $r_0$ is the bond-valence parameter which is obtained empirically.\cite{Brown1985,Brese1991} The valence for the given atom is then  

\begin{equation}
v = \sum_i{v_i} = \sum_i{\textrm{exp}[(r_0-r_i)/B]},
\label{totv}
\end{equation} 
where the sum is over all the nearest-neighbors to the atom of interest.

For YV$_4$O$_8$ and YV$_4$O$_8$, we used the bond-valence method to calculate the valences $v$ of the different inequivalent V atoms. The V atoms are bonded only to the O atoms and the V--O interatomic distances $r_i$ for the different V--O bonds at different temperatures were determined by the above Rietveld refinements of the structures of the two compounds at different temperatures. The bond-valence parameters $r_0$ for V--O bonds are listed for V$^{3+}$--O$^{2-}$, V$^{4+}$--O$^{2-}$, and V$^{5+}$--O$^{2-}$ bonds in Ref.~\onlinecite{Brown1985}. We obtained an expression for $r_0$($v_i$) by fitting the three $r_0$ versus $v_i$ values for V--O bonds\cite{Brown1985} by a second order polynomial. The valences of the four inequivalent V atoms at different temperatures for YV$_4$O$_8$ and LuV$_4$O$_8$ from Eq.~(\ref{totv}) are shown in Fig.~\ref{valency}.

\subsection{\label{mag}Magnetic measurements}
\subsubsection{\label{susc}Magnetic susceptibility}

\begin{figure*}[t]
\includegraphics[width=6in]{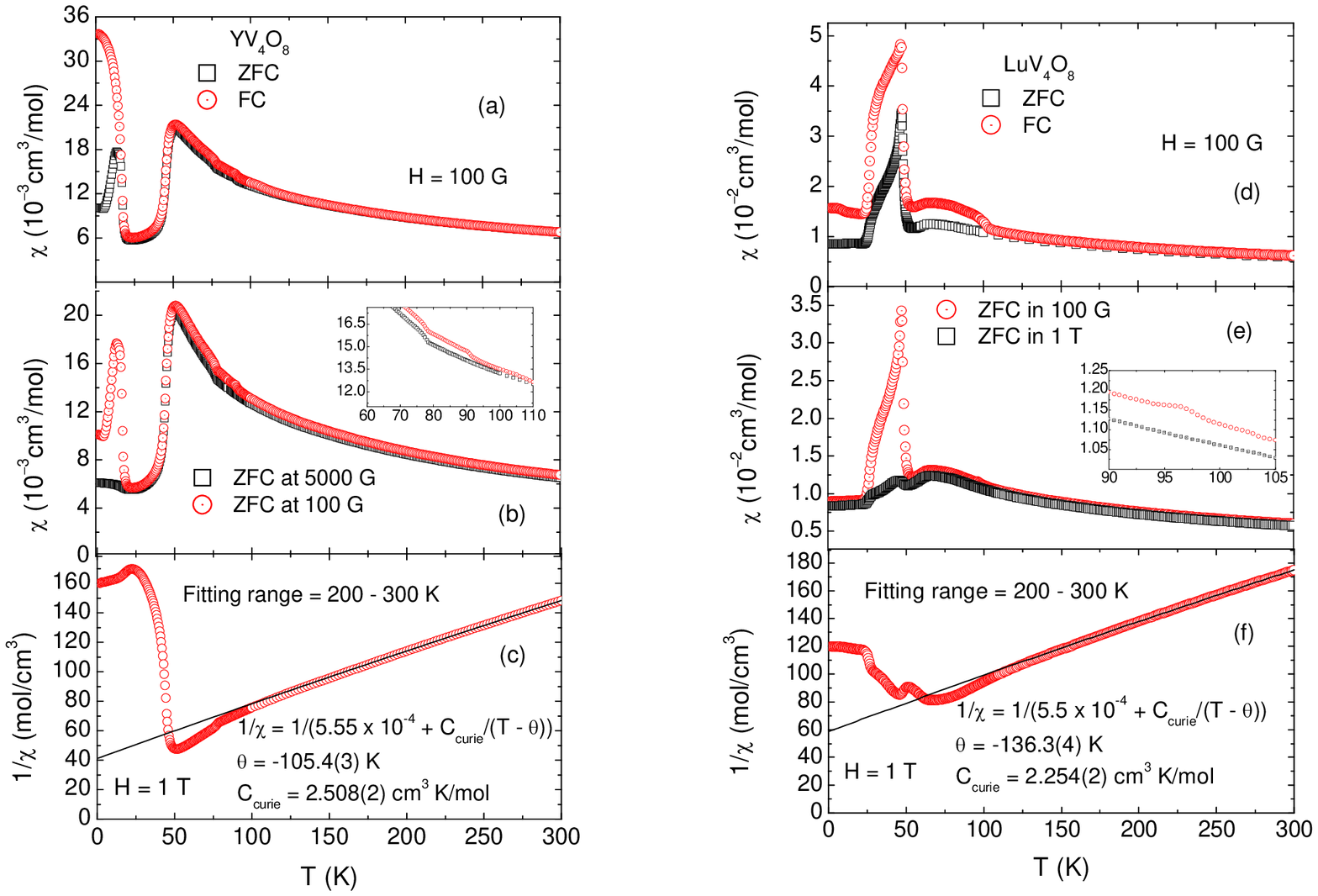}
\caption{(Color online) Zero-field-cooled (ZFC) and field-cooled (FC) magnetic susceptibility (a) ${\rm YV_4O_8}$ and (d) ${\rm LuV_4O_8}$. (b) ZFC $\chi(T)$ in 5000 G and 100 G fields of ${\rm YV_4O_8}$ and (e) ZFC $\chi$ in 100 G and 1 T fields of ${\rm LuV_4O_8}$. The insets in (b) and (e) show the the anomalies in $\chi$ at 90 K and 78 K for ${\rm YV_4O_8}$ and at 96 K for ${\rm LuV_4O_8}$, respectively. The inverse susceptibilities 1/$\chi$ versus $T$ in 1 T of ${\rm YV_4O_8}$ and ${\rm LuV_4O_8}$ are shown in (c) and (f), respectively, where the solid lines are Curie-Weiss fits to the data in the temperature range 200 -- 300 K\@.}
\label{YLususc}
\end{figure*}

\begin{table*}
\caption{Curie constant $C_{\rm Curie}$, Weiss temperature $\theta$, and temperature independent susceptibility $\chi_{\rm 0}$ of ${\rm YV_4O_8}$ and ${\rm LuV_4O_8}$ obtained from different types of Curie-Weiss fits to the inverse susceptibility 1/$\chi$ versus temperature $T$ data in the range 200 to 300 K\@. The numbers in parentheses give the error in the last digit of a quantity. The parameters which do not have errors in their values were fixed during the fittings. $\sigma^2$/DOF is the goodness of fit where $\sigma^2$ = $\sum_{i}[1/\chi(T_i) - f(T_i)]^2$ and DOF (degrees of freedom) = number of data points minus the number of fit parameters. Here $\chi(T_i)$ is the measured susceptibility $\chi$ at temperature $T$ = $T_i$ and $f(T_i)$ is the value of the fit function $f$ at $T$ = $T_i$.}
\begin{ruledtabular}
\begin{tabular}{lllll}
Compound & $\sigma^2$/DOF & $C_{\rm Curie}$ & $\chi_{\rm 0}$ & $\theta$\\
 & (10$^{-1}$ mol/cm$^3$)$^2$ & (cm$^3$ K/mol) & (10$^{-4}$ cm$^3$/mol) & (K)\\\hline
YV$_4$O$_8$ & 0.062 & 2.08(1) & 11.8(2) & $-$74(1)\\ 
 & 0.66 & 2.508(2) & 5.55 & $-$105.4(3)\\
 & 1.89 & 2.917(5) & 0 & $-$133.0(7)\\
 & 3.67 & 3.375 & $-$5.7(1) & $-$161.9(5)\\
\hline
 LuV$_4$O$_8$ & 0.12 & 1.71(1) & 12.9(2) & $-$87(1)\\ 
 & 1.39 & 2.254(2) & 5.5 & $-$136.3(4)\\
 & 3.32 & 2.698(4) & 0 & $-$172.4(7)\\
 & 5.96 & 3.375 & $-$6.78(8) & $-$216.8(5)\\
\end{tabular}
\end{ruledtabular}
\label{cwfit}
\end{table*}

Figure~\ref{YLususc}(a) shows the magnetic susceptibility $\chi \equiv M/H$ versus temperature $T$ of ${\rm YV_4O_8}$ in magnetic field $H = 100$~G\@. These data are in good agreement with the $\chi (T)$ of ${\rm YV_4O_8}$ reported in Ref.~\onlinecite{onoda03}. There is a sharp fall in the susceptibility at $T=50$~K followed by a bifurcation in the zero-field-cooled (ZFC) and field-cooled (FC) susceptibility $\chi(T)$ below 16~K\@. In addition, there are two small anomalies at $T =90$~K and $T =78$~K\@. The field dependence of $\chi$ is shown in Fig.~\ref{YLususc}(b). The sharp peak at 16~K and the small anomaly at 90~K for $H = 100$~G disappear at $H = 5000$~G\@. 

Figure~\ref{YLususc}(d) shows the ZFC and FC magnetic susceptibilities of ${\rm LuV_4O_8}$ in $H = 100$~G\@. The FC susceptibility shows a sudden slope change at $\sim 100$~K, a broad peak at $\sim 70$~K and then a sharp peak at 49~K followed by an almost $T$-independent behavior below 25~K\@. There is a strong bifurcation in the FC and ZFC susceptibility for $T<100$~K\@. The magnetic field dependence of the peak at 49~K and the small anomaly at $\sim$ 100~K are shown in Fig.~\ref{YLususc}(e). Overall, the behavior of $\chi(T)$ of YV$_4$O$_8$ and LuV$_4$O$_8$ are distinctly different.  

The high temperature $\chi(T)$ of both ${\rm YV_4O_8}$ and ${\rm LuV_4O_8}$ were fitted by the Curie-Weiss law 
\begin{equation}
\chi(T) = \chi_0 + C_{\rm Curie}/(T - \theta)\, ,
\label{cw}
\end{equation}
where $\chi_0$ is the $T$-independent magnetic susceptibility, $C_{\rm Curie}$ is the Curie constant, and $\theta$ is the Weiss temperature. The temperature range over which the data were fitted is $T = 200 - 300$ K\@.  For ${\rm YV_4O_8}$, when we let all the parameters vary, we obtained $\chi_0$ = 11.8$\times10^{-4}$~cm$^3$/mol, $C_{\rm Curie}$ = 2.08~cm$^3$K/mol, and $\theta = -74$~K\@. If we assume YV$_4$O$_8$ to be an insulator, then $\chi_0$ = $\chi_{\rm VV}$ + $\chi_{\rm dia}$ where $\chi_{\rm VV}$ is the paramagnetic Van Vleck susceptibility and $\chi_{\rm dia}$ is the diamagnetic core susceptibility. From the standard tables,\cite{shannon} we have for ${\rm YV_4O_8}$, $\chi_{\rm dia} = - 1.45\times10^{-4}$~cm$^3$/mol. The V$^{3+}$ compound ${\rm V_2O_3}$ has a $\chi_{\rm VV} \sim 2\times10^{-4}$~cm$^3$/mol~V\@.\cite{jones,takigawa} The V$^{4+}$ compound ${\rm VO_2}$ has $\chi_{\rm VV} \sim 1\times10^{-4}$~cm$^3$/mol~V\@.\cite{pouget} Thus, considering that there are three moles of V$^{3+}$ and one mole of V$^{4+}$ ions in one mole of ${\rm YV_4O_8}$, we get an estimate of $\chi_0 = 5.55\times10^{-4}$~cm$^3$/mol for ${\rm YV_4O_8}$. For ${\rm LuV_4O_8}$, we have an estimate of $\chi_0$ = 5.5$\times10^{-4}$~cm$^3$/mol. Thus, the above value of $\chi_0 = 11.8 \times 10^{-4}$~cm$^3$/mol for ${\rm YV_4O_8}$ that we obtained by fitting the data by Eq.~(\ref{cw}) with all the parameters varying is much too large. Keeping the value of $\chi_0$ fixed to 5.55$\times10^{-4}$~cm$^3$/mol, we obtain a $C_{\rm Curie}$~=~2.476(2)~cm$^3$~K/mol which is much less than the value 3.375~cm$^3$~K/mol expected for 3 V$^{3+}$ (spin $S = 1$) and 1 V$^{4+}$ ($S = 1/2$) atoms per formula unit with $g$-factor $g$ = 2. Keeping $\chi_0$ fixed to zero, we obtain a $C_{\rm Curie}$ = 2.917(5)~cm$^3$~K/mol which is closer to the expected $C_{\rm Curie}$ = 3.375 cm$^3$~K/mol. A similar analysis was done for ${\rm LuV_4O_8}$. Table~\ref{cwfit} lists the best-fit values of the parameters $C_{\rm Curie}$, $\chi_0$, and $\theta$ for ${\rm YV_4O_8}$ and LuV$_4$O$_8$ obtained in these different fits. The solid lines in Figs.~\ref{YLususc}(c) and \ref{YLususc}(f) are the Curie-Weiss fits to the 1/$\chi$ data in the temperature range 200--300 K with $\chi_0$ fixed to $5.55\times10^{-4}$~cm$^3$/mol and $5.5\times10^{-4}$~cm$^3$/mol, respectively. As shown in Figs.~\ref{YLususc}(c) and \ref{YLususc}(f), the observed inverse susceptibilities $1/\chi$ show stronger negative curvatures than the fits for both ${\rm YV_4O_8}$ and ${\rm LuV_4O_8}$. The reason might be that the temperature range of the fits is still not high enough for the Curie-Weiss law to hold. For all the fits for each compound, we see that $\theta$ is consistently negative indicating predominantly antiferromagnetic interactions between the V spins in both compounds.

\subsubsection{\label{mh}Magnetization versus applied magnetic field isotherms}

\begin{figure*}[t]
\includegraphics[width=5in]{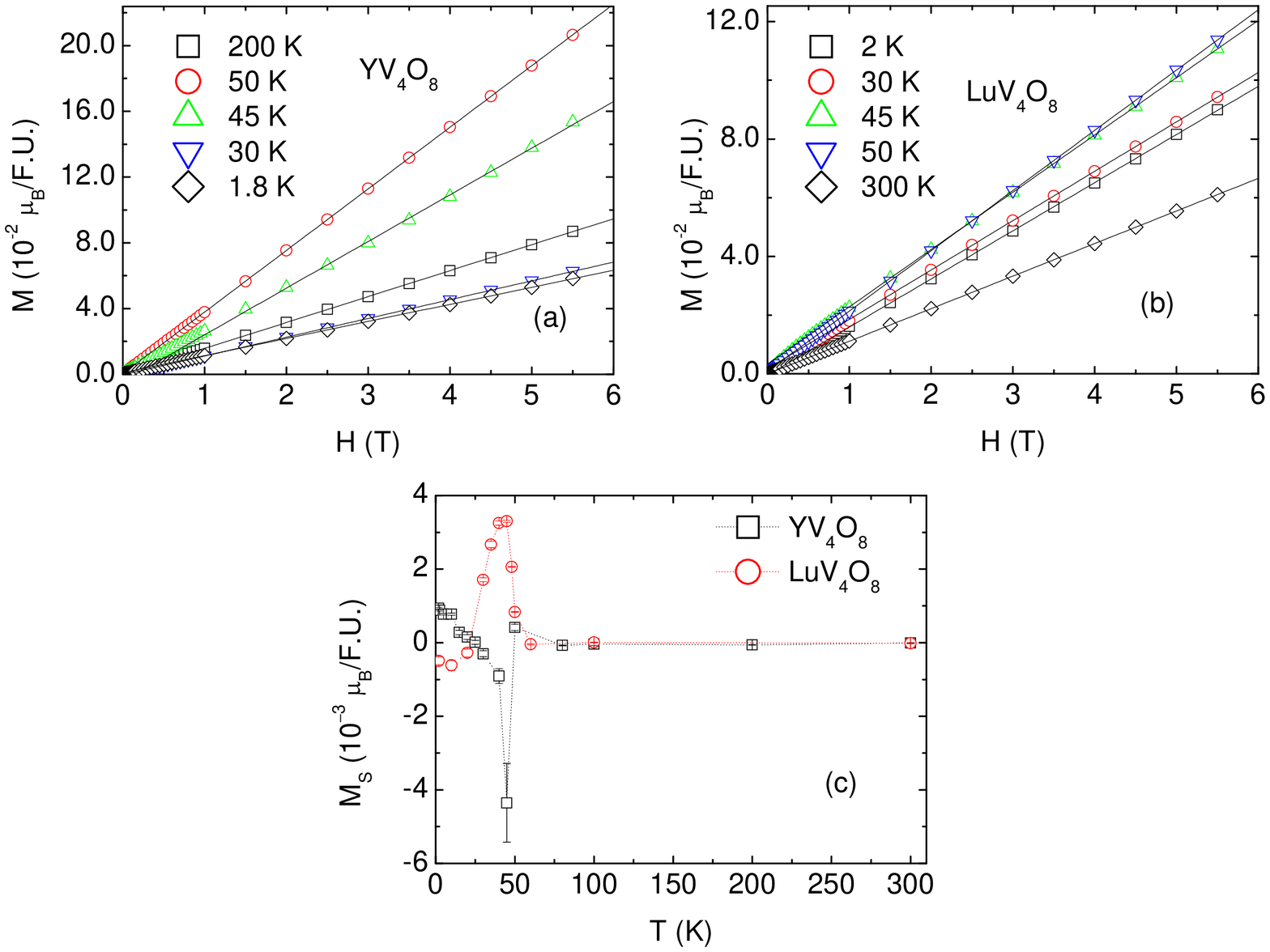}
\caption{(Color online) Magnetization $M$ versus magnetic field $H$ at different temperatures of (a) ${\rm YV_4O_8}$ and (b) ${\rm LuV_4O_8}$. The solid lines are the fits of the high field (1.5~T~$\leq~H~\leq~5.5$~T) $M(H)$ data by Eq. (\ref{fiteq}). The values of the saturation magnetization $M_{\rm S}$ versus $T$ obtained from the fits are shown in (c).}
\label{magnetization}
\end{figure*}

\begin{figure*}[t]
\includegraphics[width=5in]{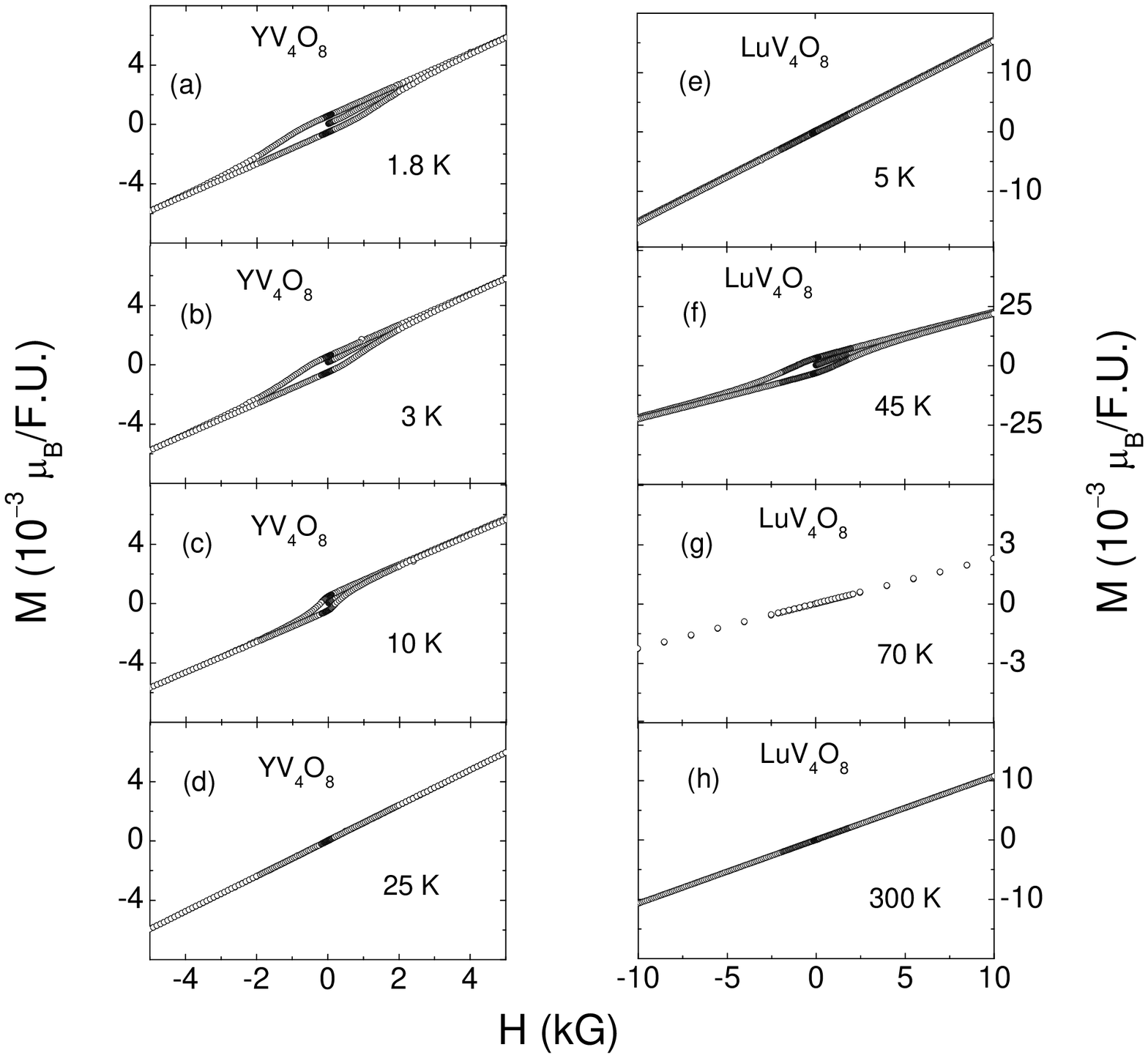}
\caption{Magnetization $M$ versus magnetic field $H$ loops at different temperatures of ${\rm YV_4O_8}$ and ${\rm LuV_4O_8}$.}
\label{hyst}
\end{figure*}

Figures~\ref{magnetization}(a) and (b) show the magnetization $M$ versus applied magnetic field $H$ isotherms at selected temperatures for LuV$_4$O$_8$ and LuV$_4$O$_8$, respectively. The saturation magnetization $\textit{M}_{\rm S}$ is obtained by fitting the high field (1.5~T~$\leq~H\leq~5.5$~T) $M(H)$ data by 
\begin{equation}
M(H,T) = \textit{M}_{\rm S}(T)\ + \chi(T) H.  
\label{fiteq}
\end{equation}
The solid lines in Figs.~\ref{magnetization}(a) and (b) are the fits of the data by Eq.~(\ref{fiteq}). The fitted $M_{\rm S}(T)$ for ${\rm YV_4O_8}$ and ${\rm LuV_4O_8}$ are shown in Fig.~\ref{magnetization}(c).  

For ${\rm YV_4O_8}$, $M_{\rm S}$ varies rapidly with temperature below 50 K\@. As temperature decreases, $M_{\rm S}$ goes to a positive value of 4.13$\times10^{-4}$ $\mu_{\rm B}$/F.U. (F.U. means formula unit) at 50 K, where $\mu_{\rm B}$ is the Bohr magneton. In view of the negative Weiss temperature found in Sec.~3 B1, this suggests a canted antiferromagnetic (AF) state. Then at 45 K, $M_{\rm S}$ decreases sharply to a negative value of 4.35$\times10^{-4}$ $\mu_{\rm B}$/F.U. which arises from an upward curvature to $M(H)$ which suggests the disappearance of canting and a sudden development of purely antiferromagnetic ordering. This is consistent with the observed susceptibility $\chi$ where $\chi$ was increasing with decreasing temperature but suddenly drops sharply at 49 K\@. As the temperature is further lowered, $M_{\rm S}$ gradually increases and finally becomes positive at 25 K and goes to a small positive value of 6.36$\times10^{-4}$ $\mu_{\rm B}$/F.U. at 1.8 K\@. 

For ${\rm LuV_4O_8}$, the behavior of $M_{\rm S}(T)$ versus $T$ is distinctly different from that of ${\rm YV_4O_8}$. As temperature decreases, $M_{\rm S}$ increases sharply from zero to 3.3$\times10^{-3}$ $\mu_{\rm B}$/F.U. at 45~K in what appears to be a first-order transition. The data suggest the development of a canted AF state below 50~K, where the canting continuously goes to zero by 20~K, which can also be observed in the susceptibility data in Fig.~\ref{YLususc}(d) where $\chi$ increases sharply at 49~K\@. Then, as the temperature is further lowered, $M_{\rm S}$ starts decreasing, becoming negative at 25~K and then remaining almost constant down to 1.8~K\@.

Figures~\ref{hyst}(a)--(d) and \ref{hyst}(e)--(h) show the $M(H)$ loops at different temperatures for ${\rm YV_4O_8}$ and ${\rm LuV_4O_8}$, respectively. For ${\rm YV_4O_8}$, measurable hysteresis is observed below 16~K\@. At 1.8~K, the remanent magnetization is 0.0007 $\mu_B$/F.U. and the coercive field is 400~G\@. For ${\rm LuV_4O_8}$, on the other hand, hysteresis is observed only around the transition at 50~K\@. At 45~K, the magnetization loop shows a remanent magnetization of 0.003 $\mu_B$/F.U. and a coercive field of 1050~G\@. As we move away from the transition at 50~K, the hysteresis disappears.

\subsection{\label{heatcap}Heat capacity measurements}

\begin{figure*}[t]
\includegraphics[width=5in]{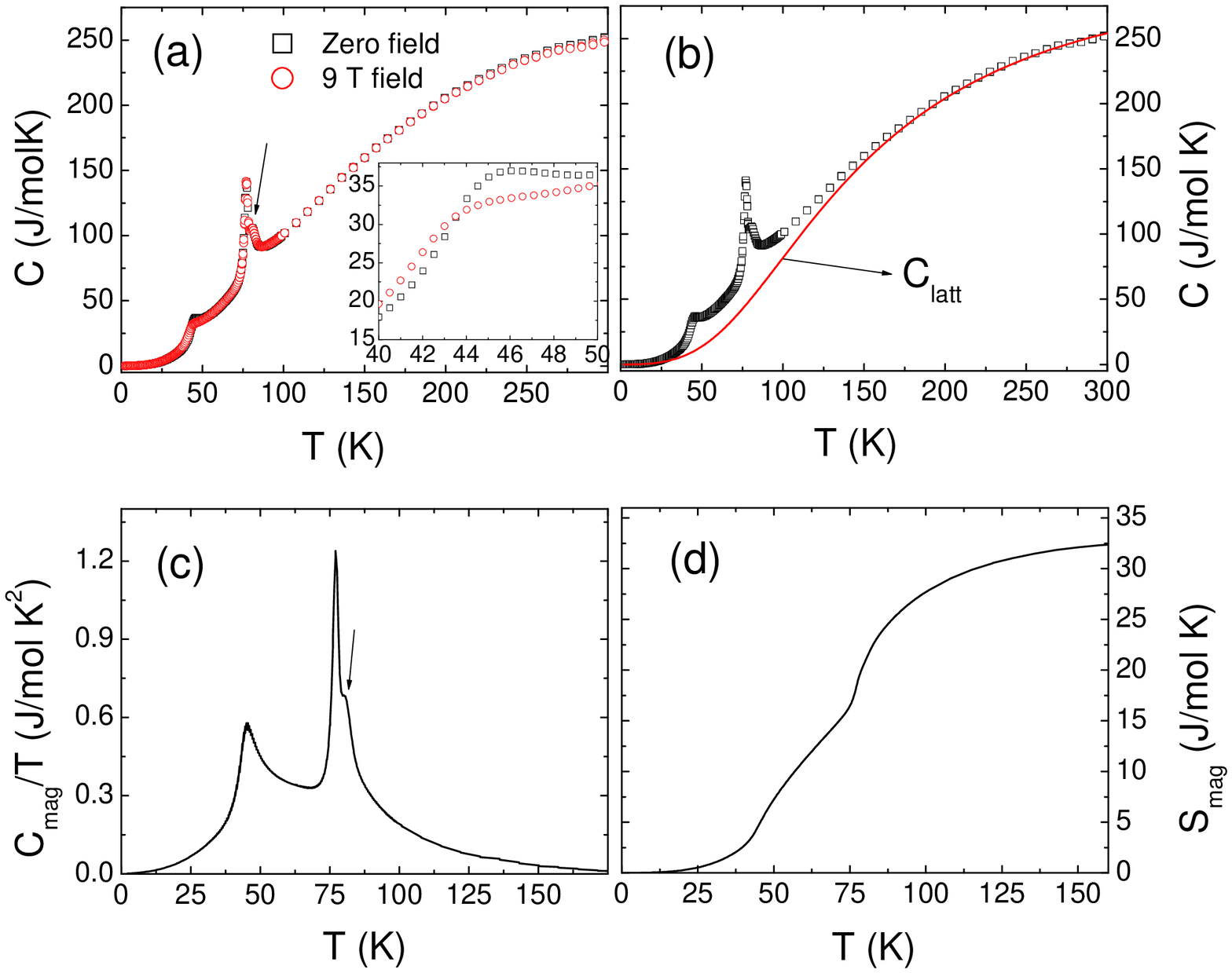}
\caption{(Color online) (a) Heat capacity $C$ versus temperature $T$ of ${\rm YV_4O_8}$ in 0 and 9 T magnetic fields. The arrow points to a tiny anomaly at 81 K\@. The inset shows a small magnetic field dependence of the heat capacity anomaly at 45 K\@. (b) The $C(T)$ in zero field from (a) along with the $C_{\rm latt}$($T$) obtained using Eq.~(\ref{debye}) with $x$ = 0.96 and $\theta_{\rm D}$ = 600 K\@. (c) $C_{\rm mag}$($T$)/$T$ versus $T$\@. The arrow points to the tiny anomaly at 81 K also seen in Fig.~\ref{hcY}(a). (d) Magnetic entropy $S_{\rm mag}(T)$ obtained from Eq.~(\ref{ent}).} 
\label{hcY}
\end{figure*}

\begin{figure*}[t]
\includegraphics[width=5in]{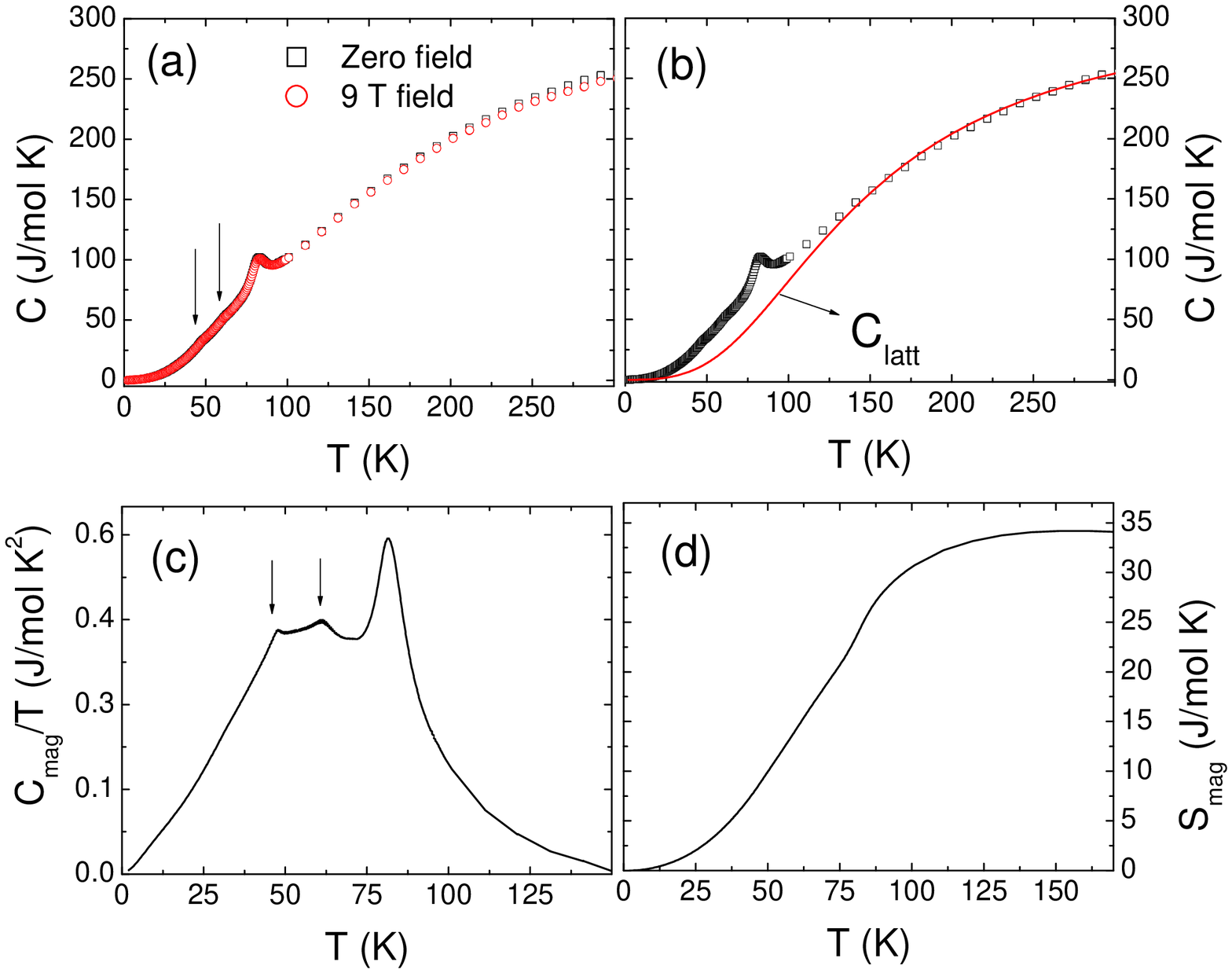}
\caption{(Color online) (a) Heat capacity $C$ versus temperature $T$ of ${\rm LuV_4O_8}$ in 0 and 9 T magnetic fields. The arrows point to two kinks at 62 K and 48 K, respectively. (b) The heat capacity $C(T)$ from (a) in zero field along with the $C_{\rm latt}$($T$) obtained from Eq.~(\ref{debye}) with $x$ = 0.96 and $\theta_{\rm D}$ = 600 K\@. (c) $C_{\rm mag}(T)$/$T$ versus $T$\@. The arrows point to the kinks at 62 K and 48 K also seen in Fig.~\ref{hcLu}(a). (d) Magnetic entropy $S_{\rm mag}(T)$ obtained from Eq.~(\ref{ent}).}
\label{hcLu}
\end{figure*}

Figure~\ref{hcY}(a) shows the molar heat capacity $C$ versus temperature $T$ of ${\rm YV_4O_8}$ in zero and 9 T magnetic fields. $C$($T$) shows a sharp peak at $T=77$ K and two small anomalies at $T=81$ K (pointed by the arrow) and $T=45$ K\@. There is a small magnetic field dependence of $C$($T$) at 45 K as shown in the inset of Fig.~\ref{hcY}(a). 

The magnetic contribution to the heat capacity $C_{\rm mag}$($T$) was obtained by $C_{\rm mag}$($T$)~=~$C(T) - C_{\rm latt}(T)$ where the lattice heat capacity $C_{\rm latt}(T)$ is estimated from the Debye model
\begin{equation}
C_{\rm latt}(T) = 9xnN_{\rm A}k_{\rm B}\Big{(}\frac{T}{\theta_{\rm D}}\Big{)}^3\int\limits_0^{\theta_{\rm D}/T}{\frac{y^4e^y}{(e^y - 1)^2}dy}\, ,
\label{debye}
\end{equation}
where $n$ is the number of atoms per formula unit, $N_{\rm A}$ is Avagadro's number, $k_{\rm B}$ is Boltzman's constant, $\theta_{\rm D}$ is the Debye temperature, and $x$ is a scaling factor which we had to introduce to get a considerable overlap of Eq.~(\ref{debye}) with the measured $C$\@ at high $T$. Plots of $C_{\rm latt}$ versus $T$ were obtained for various values of the Debye temperature $\theta_{\rm D}$ and $x$, and were compared to the plot of measured $C(T)$ versus $T$\@. The $C_{\rm latt}(T)$ with the maximum overlap with the plot of $C(T)$ data at high temperatures was chosen. 

For ${\rm YV_4O_8}$, we obtained the best fit of $C_{\rm latt}$($T$) by Eq.~(\ref{debye}) with $\theta_{\rm D}$ = 600 K and $x$ = 0.96 for $T > 200$~K\@. Figure~\ref{hcY}(b) shows the plot of $C_{\rm latt}(T)$ along with the measured $C(T)$ for ${\rm YV_4O_8}$. Figure~\ref{hcY}(c) shows the magnetic contribution to the heat capacity $C_{\rm mag}(T)$/$T$ $\equiv$ [$C(T) - C_{\rm latt}(T)$]/$T$ for ${\rm YV_4O_8}$ and Fig.~\ref{hcY}(d) shows the magnetic entropy $S_{\rm mag}(T)$ versus $T$ of ${\rm YV_4O_8}$ given by 
\begin{equation}
S_{\rm mag}(T) = \int\limits_{0}^{T}{\frac{C_{\rm mag}(T)}{T}dT}\, .
\label{ent}
\end{equation}
The change in $S_{\rm mag}$ over the temperature range 0 K to 90 K in which the magnetic transitions occur is 32.5 J/mol K\@. If the V spins order, then the magnetic entropy associated with the spin ordering $S_{\rm spin}$ is given by
\begin{equation}
S_{\rm spin} = \sum_{i}{n_iR{\rm ln}(2S_i+1)}\, ,
\label{spinentropy}
\end{equation}
where the sum is over V spins $S_i$ in a formula unit, $n_i$ is the number of spins $S_i$, and $R$ is the molar gas constant. Using $n_i$ = 3 V$^{+3}$ ($S$ = 1) and 1 V$^{+4}$ ($S$ = 1/2) per formula unit gives $S_{\rm mag}$ = 33.14~J/mol K which is very close (within 2\%) to the value of $S_{\rm mag}$ obtained above. This indicates that our estimation of $C_{\rm latt}$($T$) is reasonable.

Figure~\ref{hcLu}(a) shows the $C$($T$) of ${\rm LuV_4O_8}$ in zero and 9 T magnetic fields. There is a peak at $T=80$ K and two small kinks at 62 K and 48 K, pointed out by two arrows, respectively. The magnetic field dependence of $C(T)$ is negligible. Figure~\ref{hcLu}(b) shows the zero field $C$($T$) and the $C_{\rm latt}$($T$) for ${\rm LuV_4O_8}$ from Eq.~(\ref{debye}). For ${\rm LuV_4O_8}$, the values $\theta_{\rm D}$ = 600 K and $x$ = 0.96 produced the $C_{\rm latt}$($T$) with the maximum overlap with $C$($T$) at high $T > 150$ K\@. Figure~\ref{hcLu}(c) shows $C_{\rm mag}$($T$)/$T$ versus $T$ for ${\rm LuV_4O_8}$. The two kinks pointed out by the arrows in Fig.~\ref{hcLu}(a) can be seen prominently here. The magnetic entropy $S_{\rm mag}$ calculated from Eq.~(\ref{ent}) versus $T$ is shown in Fig.~\ref{hcLu}(d). The total magnetic entropy change up to 150~K is 34.0 J/mol K, which again agrees very well with the the above value of 33.1 J/mol K for disordered V spins. A sharp peak occurs in $C_{\rm mag}(T)$ at $\approx 80$~K with two additional kinks highlighted by two vertical arrows at 45~K and $\approx 60$~K, respectively, as shown in Fig.~\ref{hcLu}(c).

\subsection{\label{transport}Electrical resistivity measurements}

\begin{figure*}[t]
\includegraphics[width=5in]{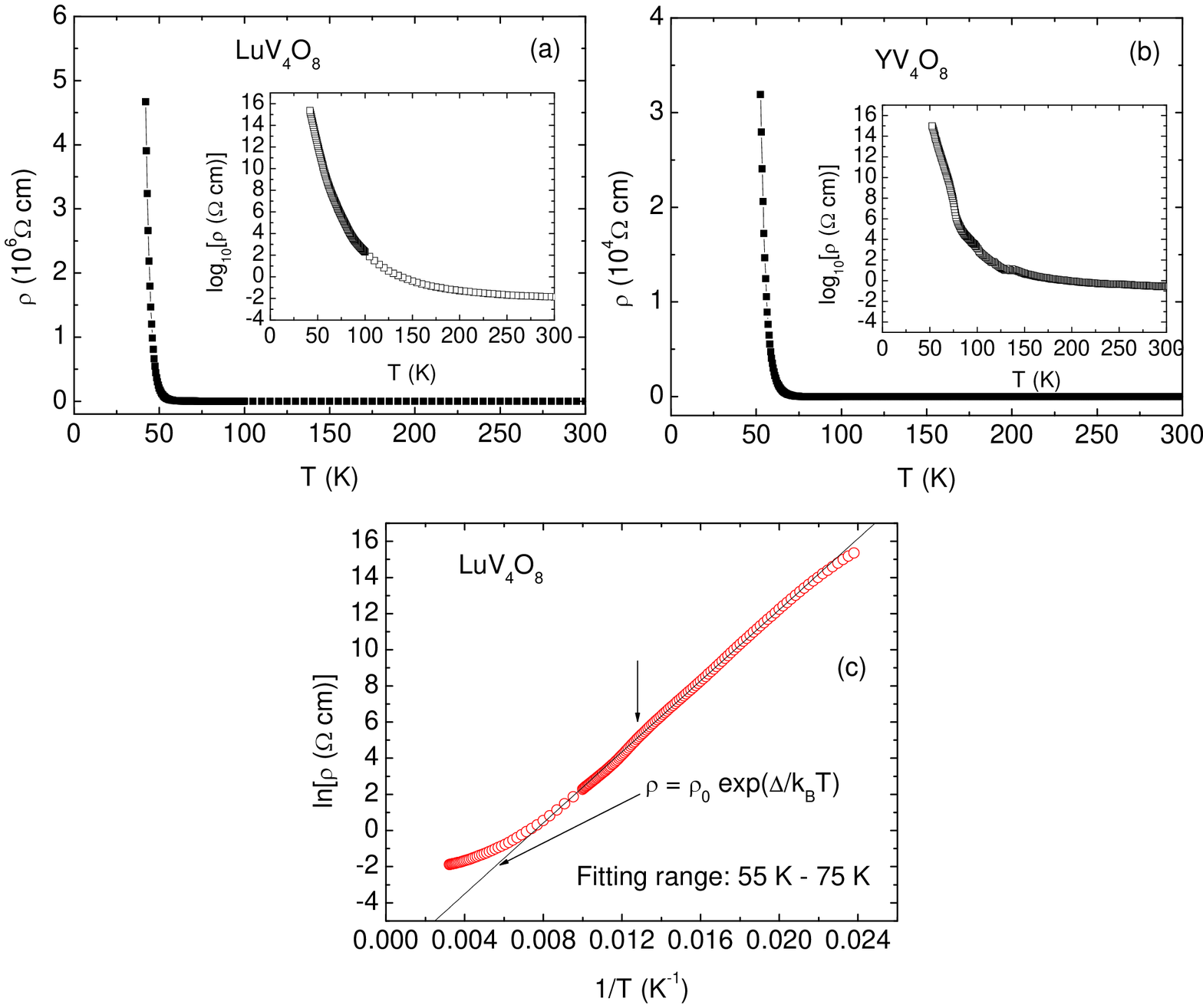}
\caption{Electrical resistivity $\rho$ versus temperature $T$ measured on sintered pellets of (a) ${\rm LuV_4O_8}$ and (b) ${\rm YV_4O_8}$. Insets in (a) and (b) show log($\rho$) versus $T$ for ${\rm LuV_4O_8}$ and ${\rm YV_4O_8}$, respectively. (c) ln($\rho$) versus $1/T$ for LuV$_4$O$_8$. The solid line in (c) is the fit to the data by Eq.~(\ref{activation}) in the temperature range 55~K ($1/T = 0.018$~K) to 75~K ($1/T = 0.0133$~K) where the data are approximately linear.}
\label{res}
\end{figure*}

Figures~\ref{res}(a) and (b) show the electrical resistivity $\rho$ versus temperature $T$ measured on  pieces of sintered pellets of LuV$_4$O$_8$ and YV$_4$O$_8$, respectively. On the scale of the figures, the resistivities are nearly temperature-independent above 50~K and 60~K, respectively, and strongly increase below those temperatures, suggesting the occurrence of metal to insulator transitions upon cooling below those temperatures. The insets in Figs.~\ref{res}(a) and (b) show the respective log$_{10}$($\rho$) versus $T$ for the two compounds. For both compounds, log$_{10}$($\rho$) increases with decreasing $T$ showing apparent semiconducting behaviors over the whole $T$ range. However, the nearly $T$-independent behaviors at the highest temperatures suggest metallic behavior as just noted. Polycrystalline pellets of metallic oxides are notorious for showing semiconducting-like behavior due to insulating material in the grain boundaries. A plot of ln($\rho$) versus $1/T$ for LuV$_4$O$_8$ is shown in Fig.~\ref{res}(c).  We fitted these data by 
\begin{equation}
\rho = \rho_0 {\rm exp}[\Delta/k_{\rm B}T],
\label{activation}
\end{equation} 
where $\Delta$ is the activation energy, $\rho_0$ is a constant, and $k_{\rm B}$ is Boltzmann's constant. The solid line in Fig.~\ref{res}(c) is the fit in the $T$ range 55~K ($1/T = 0.018$ K) to 75 K ($1/T = 0.0133$ K) where the ln[$\rho(1/T)$] data are approximately linear. The obtained fit parameters are $\rho_0 = 7.44(3)$~$\Omega$~cm and $\Delta = 84.6(1)$~meV\@.

\section{\label{discussion}Discussion}

\begin{table}
\caption{Different temperatures at which anomalies were observed in $\chi(T)$, $C(T)$, and $\rho(T)$, respectively, for YV$_4$O$_8$.}
\begin{ruledtabular}
\begin{tabular}{llll}
& $\chi$ & $C$ & $\rho$\\
\hline
$T_1$ & 16 & &\\
$T_2$ & 50 &45 &\\
$T_3$ &  & &60\\
$T_4$ & 78 & 77 &\\
$T_5$ & & 81 &\\
$T_6$ & 90 & &\\
\end{tabular}
\end{ruledtabular}
\label{tabfinalY}
\end{table}

\begin{table}
\caption{Different temperatures at which anomalies were observed in $\chi(T)$, $C(T)$, and $\rho(T)$, respectively, for LuV$_4$O$_8$.}
\begin{ruledtabular}
\begin{tabular}{llll}
& $\chi$ & $C$ & $\rho$\\
\hline
$T_1$ & 25 & &\\
$T_2$ & 50 &48 &50\\
$T_3$ &  & 62 &\\
$T_4$ &  & 80 &\\
$T_5$ & 100 & &\\
\end{tabular}
\end{ruledtabular}
\label{tabfinalLu}
\end{table}

Tables~\ref{tabfinalY} and \ref{tabfinalLu} list the temperatures at which anomalies were observed in the $\chi(T)$, $C(T)$, and $\rho(T)$ measurements of YV$_4$O$_8$ and LuV$_4$O$_8$, respectively. Upon cooling below $\approx$~50~K, a sharp decrease of the V1V31 distance, increase of the V1V32 distance, and an increase in the other V1-V3 distances as shown in Fig.~\ref{VVdis}(b) suggest dimerization of the V1 and V3 spins in the V1-V3 chain (see Fig.~\ref{VVdisstruc}) in YV$_4$O$_8$. The valences of V1 and V3 from Fig.~\ref{valency}(a) are close to 3 suggesting that both have spin $S = 1$. From the Curie-Weiss fit of the magnetic susceptibility in Fig.~\ref{YLususc}(c), the dominant interactions between the V spins are antiferromagnetic. We infer that the dimerization leads to a suppression of the magnetic susceptibility in the V1-V3 chain below 50 K\@. For the other V2-V4 chain, below 50 K, all the V-V interatomic distances increase as shown in Fig.~\ref{VVdis}(a), allowing the spins to order antiferromagnetically. The calculated valences of the V2 and V4 atoms in Fig.~\ref{valency}(a) point towards a decrease in the spin states of those V atoms. Both effects probably contribute to the sudden sharp drop in the magnetic susceptibility below 50 K in Figs.~\ref{YLususc}(a) and (b).

The transition observed in $\chi(T)$ at 50 K in Figs.~\ref{YLususc}(a) and (b) for YV$_4$O$_8$ also appears in $C_{\rm mag}(T)$ and $\rho(T)$ for this compound at a similar temperature in Figs.~\ref{hcY}(a), (c), and Fig.~\ref{res}(b). The presence of the anomaly in $C(T)$ strengthens our interpretation of dimerization due to structural transition and long range antiferromagnetic ordering at 50 K\@. However, there is no anomaly in $C_{\rm mag}$ at 16 K where the ZFC-FC $\chi(T)$ data in Fig.~\ref{YLususc}(a) show a strong bifurcation which disappears at high fields as shown in Fig.~\ref{YLususc}(b). No change in $C_{\rm mag}(T)$ in Fig.~\ref{hcY}(c) is observed at 16~K, suggesting that the bifurcation of the ZFC-FC $\chi(T)$ may be due to weak canting of the antiferromagnetically ordered V spins. The presence of magnetic hysteresis with a very small (0.0007 $\mu_{\rm B}$/F.U.) remnant magnetization at 2 K shown in Fig.~\ref{hyst}(a) and a small almost $T$-independent $M_{\rm S}(T)$ below 16~K shown in Fig.~\ref{magnetization}(c) are all consistent with the occurrence of canted antiferromagnetism below 16~K\@. There are two additional anomalies at 75~K and 90~K which appear in both $\chi(T)$ and $C_{\rm mag}(T)$, the origins of which are unclear.

The dimerization of the V spins in one of the chains and formation of spin singlets in YV$_4$O$_8$ is very similar to the spin-Peierls transition observed in CuGeO$_3$ at 14 K\@.\cite{masashi} The occurrence of a metal to insulator transition at 60~K (which is very close to the temperature of the spin singlet formation) as shown in Fig.~\ref{res}(b) suggests that YV$_4$O$_8$ is a rare example where a metal to spin singlet insulator transition takes place. Such a Peierls-like transition has been observed in the tetragonal rutile VO$_2$ at 340 K\cite{morin1959,pouget1974} and in the spinel MgTi$_2$O$_4$ at 260~K\@.\cite{masahiko,zhou} In both VO$_2$ and MgTi$_2$O$_4$, a complete structural transition occurs at the temperature of the metal to spin singlet transition,\cite{goodenough1971,masahiko} unlike YV$_4$O$_8$, where only the lattice parameters change without a lowering of the crystal symmetry.     

For LuV$_4$O$_8$, the magnetic susceptibility in Figs.~\ref{YLususc}(d) and (e) shows no evidence of formation of spin singlets. There is no anomaly in $C_{\rm mag}(T)$ in Fig.~\ref{hcLu}(c) at $\approx 100$~K at which a slope change occurs in $\chi(T)$ in Fig.~\ref{YLususc}(d). On the other hand, a sharp peak occurs in $C_{\rm mag}(T)$ at $\approx 80$~K, where no anomaly in $\chi(T)$ occurs. This might indicate the onset of short-range ordering at $\approx 100$~K followed by long-range ordering at $\approx 80$~K\@. From Figs.~\ref{YLususc}(d) and (e), the $\chi(T)$ shows a sharp increase at $\approx 50$~K, whereas in Fig.~\ref{hcLu}(c) there is only a small kink in $C_{\rm mag}(T)$ at this $T$\@. The absence of a sharp anomaly in $C_{\rm mag}$ at 50 K might indicate the development of a canted AF state at that temperature.

The Curie-Weiss fits to the high $T$ $\chi$ for both YV$_4$O$_8$ and YV$_4$O$_8$ yield Curie constants that are considerably lower than expected, which leads to the possibility of both these compounds being metallic.

\section{\label{summary}Summary}

We have synthesized powder samples of YV$_4$O$_8$ and LuV$_4$O$_8$ whose crystallographic structure consists of two distinct one-dimensional zigzag chains running along the crystallographic $c$-axis. X-ray diffraction measurements down to 10 K reveal a  first-order-like phase transition with a sudden change in the lattice parameters and unit cell volume at 50 K in YV$_4$O$_8$. However, the high and low temperature structures could be refined using the same space group indicating no lowering of the symmetry of the unit cell due to the structural transition. As a result of the transition, one of the chains dimerizes. The magnetic susceptibility of YV$_4$O$_8$ exhibits a sharp first-order-like decrease at 50 K followed by a bifurcation in the ZFC-FC susecptibility below 16 K\@. The anomaly at 50 K is suggested to arise from the dimerization of the $S = 1$ chain and antiferromagnetic (AF) ordering of the other chain. The AF ordered spins then become canted below 16 K\@. The change in the magnetic entropy calculated from heat capacity measurements also agrees very well with ordering of three $S = 1$ and one $S = 1/2$ disordered spins per formula unit. The lattice parameters of LuV$_4$O$_8$ exhibit a small anomaly at $\sim 50$ K but not as sharp as in YV$_4$O$_8$. The magnetic susceptibility of LuV$_4$O$_8$ shows a broad peak at $\sim 60$ K followed by a sharp first order-like increase at 50 K\@. The 50 K anomaly is suppressed at higher fields. For both compounds, Curie-Weiss fits to the high $T$ susceptibilities yield Curie constants which are much lower than expected. Electrical resistivity measurements on sintered pellets indicate metal to insulator-like transition at 60~K and 50~K for YV$_4$O$_8$ and LuV$_4$O$_8$, respectively. It would be very interesting to study single crystals of these compounds. Single crystal resistivity measurements are needed to determine if these materials are metallic or not at high temperatures. Measurements such as NMR or neutron scattering that would provide microscopic information about the spin dynamics would also be valuable to clarify the nature of the magnetic ordering transitions in ${\rm YV_4O_8}$ and ${\rm LuV_4O_8}$.

\begin{acknowledgments}
Work at the Ames Laboratory was supported by the Department of Energy-Basic Energy Sciences under Contract No. DE-AC02-07CH11358.
\end{acknowledgments}


\begin{thebibliography} {30}

\bibitem{udo2004} U. Schwingenschl$\ddot{\rm o}$gl and V. Eyert, Ann.\ Phys.\ (Leipzig) \textbf{13}, 475 (2004).

\bibitem{kachi73} S. Kachi, K. Kosuge, and H. Okinaka, J. Solid State Chem.\ \textbf{6}, 258 (1973).

\bibitem{kondo97} S. Kondo, D. C. Johnston, C. A. Swenson, F. Borsa, A. V. Mahajan, L. L. Miller, T. Gu, A. I. Goldman, M. B. Maple, D. A. Gajewski, E. J. Freeman, N. R. Dilley, R. P. Dickey, J. Merrin, K. Kojima, G. M. Luke, Y. J. Uemura, O.~Chmaissem, and J. D. Jorgensen, Phys.\ Rev.\ Lett.\ \textbf{78}, 3729 (1997).

\bibitem{johnston77} D. C. Johnston, J.\ Low Temp.\ Phys.\ \textbf{255}, 145 (1976).

\bibitem{decker} B. F. Decker and J. S. Kasper, Acta Crystallogr.\ \textbf{10}, 332(1957).

\bibitem{hastings} J. M. Hastings, L. M. Corliss, W. Kunnmann, and S. La Placa, J.\ Phys.\ Chem.\ Solids \textbf{28}, 1089(1967).


\bibitem{niazi2009} A. Niazi, S. L. Bud'ko, D. L. Schlagel, J. Q. Yan, T. A. Lograsso, A. Kreyssig, S. Das, S. Nandi, A. I. Goldman, A. Honecker, R. W. McCallum, M. Reehuis, O. Pieper, B. Lake, and D. C. Johnston, Phys.\ Rev.\ B \textbf{79}, 104432 (2009).

\bibitem{zong2008} X. Zong, B. J. Suh, A. Niazi, J. Q. Yan, D. L. Schlagel, T. A. Lograsso, and D. C. Johnston, Phys.\ Rev.\ B \textbf{77}, 014412 (2008).


\bibitem{pieper2009} O. Pieper, B. Lake, A. Daoud-Aladine, M. Reehuis, K. Prokes, B. Klemke, K. Kiefer, J. Q. Yan, A. Niazi, D. C. Johnston, and A. Honecker, Phys.\ Rev.\ B \textbf{79}, 180409(R) (2009). 

\bibitem{hikihara} T. Hikihara, M. Kaburagi, H. Kawamura, and T. Tonegawa, J.\ Phys.\ Soc.\ Jpn.\ \textbf{69}, 259 (2000).

\bibitem{Yamaura2007} K. Yamaura, M. Arai, A. Sato, A. B. Karki, D. P. Young, R. Movshovich, S. Okamoto, D. Mandrus, and E. Takayama-Muromachi, Phys.\ Rev.\ Lett.\ \textbf{99}, 196601 (2007).

\bibitem{Sakurai2008} H. Sakurai, Phys.\ Rev.\ B\ \textbf{78}, 094410 (2008).

\bibitem{kanke97} Y. Kanke and K. Kato, Chem.\ Mater.\ \textbf{9}, 141 (1997).

\bibitem{friese07} K. Friese, Y. Kanke, A. N. Fitch, and A. Grzechnik, Chem.\ Mater.\ \textbf{19}, 4882 (2007).

\bibitem{onoda03} M. Onoda, Acta.\ Cryst.\ \textbf{B59}, 429 (2003).

\bibitem{kitayama78} K. Kitayama, Bull.\ Chem.\ Soc.\ Jpn.\ \textbf{51}, 1358 (1978).


\bibitem{bystrom} A. Bystr$\ddot{\rm o}$m and A. M. Bystr$\ddot{\rm o}$m, Acta Cryst.\ \textbf{3}, 146 (1950).

\bibitem{torardi1985} C. C. Torardi, Mater.\ Res.\ Bull.\ \textbf{20}, 705 (1985).

\bibitem{isobe} M. Isobe, S. Koishi, N. Kouno, J.-I. Yamaura,
T. Yamauchi, H. Ueda, H. Gotou, T. Yagi, and Y. Ueda, J.\ Phys.\ Soc.\ Jpn.\ \textbf{75}, 073801 (2006).

\bibitem{cava2003} Z. Q. Mao, T. He, M. M. Rosario, K. D. Nelson, D. Okuno, B. Ueland, I. G. Deac, P. Schiffer, Y. Liu, and R. J. Cava, Phys.\ Rev.\ Lett.\ \textbf{90}, 186601 (2003).




\bibitem{gsas} A. C. Larson and R. B. Von Dreele, ``General Structure Analysis System (GSAS)'', Los Alamos National Laboratory Report LAUR 86-748 (2000); B. H. Toby, J.\ Appl.\ Cryst.\ \textbf{34}, 210 (2001). 

\bibitem{Brown1973} I. D. Brown and R. D. Shannon, Acta Cryst.\ Sect.\ A \textbf{29}, 266 (1973).

\bibitem{Brown1985} I. D. Brown and D. Altermatt, Acta Cryst.\ \textbf{B41}, 244 (1985). 

\bibitem{Brese1991} N. E. Brese and M. O'Keeffe, Acta Cryst.\ \textbf{B47}, 192 (1991).

\bibitem{shannon} R. D. Shannon, Acta Crystallogr.\ \textbf{A32}, 751 (1976).

\bibitem{jones} E. D. Jones, Phys.\ Rev.\ \textbf{137}, A978 (1965).

\bibitem{takigawa} M. Takigawa, E. T. Ahrens, and Y. Ueda, Phys.\ Rev.\ Lett.\ \textbf{76}, 283 (1996).

\bibitem{pouget} J. P. Pouget, D. S. Schreiber, H. Launois, D. Wohlleben, A. Casalot, and G. Villeneuve, J.\ Phys.\ Chem.\ Solids \textbf{33}, 1961 (1972).

\bibitem{holm} A. P. Holm, V. K. Pecharsky, K. A. Gschneidner, Jr., R. Rink, and M. Jirmanus, Rev.\ Sci.\ Instrum.\ \textbf{75}, 1081 (2004).

\bibitem{masashi} M. Hase, I. Terasaki, and K. Uchinokura, Phys.\ Rev.\ Lett.\ \textbf{70}, 3651 (1993).

\bibitem{morin1959} F. J. Morin, Phys.\ Rev.\ Lett.\ \textbf{3}, 34 (1959).

\bibitem{pouget1974} J. P. Pouget, H. Launois, T. M. Rice, P. Dernier, A. Gossard, G. Villeneuve, and P. Hagenmuller, Phys.\ Rev.\ B\ \textbf{10}, 1801 (1974).  

\bibitem{masahiko} M. Isobe and Y. Ueda, J.\ Phys.\ Soc.\ Jpn.\ \textbf{71}, 1848 (2002).

\bibitem{zhou} J. Zhou, G. Li, J. L. Luo, Y. C. Ma, Dan Wu, B. P. Zhu, Z. Tang, J. Shi, and N. L. Wang, Phys.\ Rev.\ B\ \textbf{74}, 245102 (2006).

\bibitem{goodenough1971} J. B. Goodenough, J.\ Solid State Chem.\ \textbf{3}, 490 (1971).

\end{thebibliography}
\end{document}